\documentclass[aps,twocolumn,epsfig,superscriptaddress,showpacs]{revtex4}
\usepackage{graphicx} 
\usepackage{psfrag} 
\usepackage{filecontents}
\usepackage{color}
\usepackage{bm}

\usepackage[
colorlinks=true,
citecolor=blue,
urlcolor=blue,
setpagesize=false]{hyperref}

\begin{document}

\title{Electronic structures of iMAX phases and their two-dimensional derivatives: 
A family of piezoelectric materials}

\author{Mohammad Khazaei}
\email{khazaei@riken.jp}
\affiliation{Computational Materials Science Research Team, RIKEN Center for Computational Science (RCCS), Kobe, Hyogo 650-0047, Japan}
\author{Vei Wang\footnote{V.W. and C.S. contributed equally to this work.}}
\affiliation{Department  of  Applied  Physics, Xi'an University of Technology, Xi'an 710054, China}
\author{Cem Sevik}
\affiliation{Department of Mechanical Engineering, Anadolu University, Eskisehir 26555, Turkey}
\author{Ahmad Ranjbar}
\affiliation{Computational Materials Science Research Team, RIKEN Center for Computational Science (RCCS), Kobe, Hyogo 650-0047, Japan}
\author{Masao Arai}
\affiliation{International Center for Materials Nanoarchitectonics, National Institute for Materials Science (NIMS), 1-1 Namiki, 
Tsukuba 305-0044, Ibaraki, Japan}
\author{Seiji Yunoki}
\affiliation{Computational Materials Science Research Team, RIKEN Center for Computational Science (RCCS), Kobe, Hyogo 650-0047, Japan}
\affiliation{Computational Condensed Matter Physics Laboratory, RIKEN, Wako, Saitama 351-0198, Japan}
\affiliation{Computational Quantum Matter Research Team, RIKEN Center for Emergent Matter Science (CEMS), Wako, Saitama 351-0198, Japan}

\date{\today}

\begin{abstract}

Recently, a group of MAX phases, (Mo$_{2/3}$Y$_{1/3}$)$_2$AlC, (Mo$_{2/3}$Sc$_{1/3}$)$_2$AlC, (W$_{2/3}$Sc$_{1/3}$)$_2$AlC, (W$_{2/3}$Y$_{1/3}$)$_2$AlC, and 
(V$_{2/3}$Zr$_{1/3}$)$_2$AlC, with in-plane ordered double transition metals, named iMAX phases, 
have been synthesized. 
Experimentally, some of these MAX phases can be chemically exfoliated into two-dimensional (2D) 
single- or multilayered transition metal carbides, so-called MXenes. 
Accordingly, the  2D nanostructures derived from iMAX phases are named iMXenes. 
Here, we investigate the structural stabilities and electronic structures of the experimentally discovered 
iMAX phases and their possible iMXene derivatives. 
We show that the iMAX phases and their pristine, F, or OH-terminated iMXenes are metallic. 
However, upon O termination, (Mo$_{2/3}$Y$_{1/3}$)$_2$C, (Mo$_{2/3}$Sc$_{1/3}$)$_2$C, (W$_{2/3}$Y$_{1/3}$)$_2$C, and (W$_{2/3}$Sc$_{1/3}$)$_2$C iMXenes 
turn into semiconductors.  
The semiconducting iMXenes may find applications in piezoelectricity owing to the absence of centrosymmetry.
Our calculations reveal that the semiconducting iMXenes possess giant piezoelectric coefficients as large as 
45$\times10^{-10}$~C/m.
\end{abstract}


\maketitle


\section{Introduction}
Over the past couple of years, tremendous improvements have been achieved in the synthesis of crystalline MAX phases \textemdash a large family of transition metal 
carbides and nitrides with layered hexagonal structures and a chemical formula of M$_{n+1}$AX$_{n}$, where ``M" is an early transition metal (Sc, Ti, V, Nb, Ta, Cr, and Mo), 
``A" belongs to groups \textrm{XIII}$-$\textrm{XVI} in the periodic table 
(Al, Ga, In, Tl, Si, Ge, Sn, P, As, Bi, and S), ``X" stands for C or N, and 
$n$=1$-$3~\cite{M.W.Barsoum2000,J.Wang2009,Z.M.Sun2011,D.Horlait2016,H.Fashandi2017}. Traditionally, MAX phases are well-known materials because of 
their ceramic nature with a high structural stability against fatigue, creep, thermal shock, and corrosive reactions; 
therefore, they have been adopted to many 
structural applications~\cite{M.W.Barsoum2000,J.Wang2009}. In recent experiments, it has been shown that the ``A'' element can be removed from some MAX phases 
by using combinations of an appropriate acid treatment and 
sonication. This process  results in the formation of single or multilayered two-dimensional (2D) transition metal carbides or nitrides,
so-called MXenes~\cite{M.Naguib2011,M.Naguib2012}. During the exfoliation process, depending on the type of acid solution (e.g., HF acid), the 
surfaces of MXenes are usually terminated with a mixture of F, OH, and O groups~\cite{M.Naguib2011,M.Naguib2012,I.Persson2017}. 
As stated in recent reviews covering the current status of research and development 
on this novel family of low-dimensional systems~\cite{B.Anasori2017_1,M.Khazaei2017}, 
MXenes may have potential 
electronic~\cite{M.Khazaei2013,M.Khazaei2014_3,M.Khazaei2015,H.Weng2015,S.Lai2015,M.Khazaei2016_1,M.Khazaei2016_2,F.Shahzad2016,Y. Liang2017,Y.Yang2017},
magnetic~\cite{M.Khazaei2013,C.Si2015,J.He2016,M.Je2016,G.Gao2016,H.Kumar2017,T.Ouisse2017,Y.Zhang2017}, 
thermoelectric~\cite{M.Khazaei2013,M.Khazaei2014_3}, 
optoelectronic~\cite{H.Lashgari2014}, 
ferroelectric~\cite{A.Chandrasekaran2017},
superconducting~\cite{F.Zhang2017,J.J.Zhang2017}, and 
energy storage applications~\cite{Y.Xie2014_1,R.B.Rakhi2015,H.Zhang2017}, 
as well as applications in energy conversion devices~\cite{H.Lin2016,H.Kim2017,G.Fan2017,J.Ran2017}  
and in developing new composite materials~\cite{X.Zhang2013,M.Xue2017}. 
Hence, interest in the synthesis of novel MAX phases and 2D MXenes has 
rapidly increased. 

\begin{figure}[t]
\centering
  \includegraphics[scale=0.235]{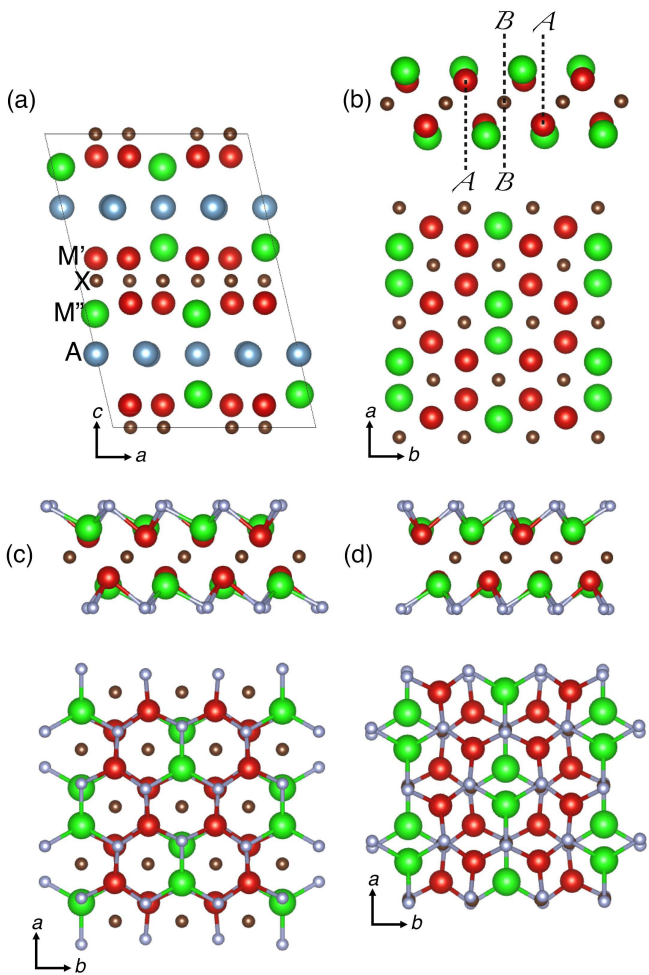}
  \caption{
  (a) Side view of the conventional cell (indicated by thin lines)
  of (M$'_{2/3}$M$''_{1/3}$)$_2$AX iMAX phase. 
  (b) Side and top views of pristine (M$'_{2/3}$M$''_{1/3}$)$_2$X iMXenes. $\cal{A}$ and $\cal{B}$ indicate different sites on the surfaces of iMXenes
  for adsorption of chemical groups. 
 (c) Side and top views of $\cal{AA}$ model of functionalized iMXenes with chemical groups (F, O, or OH). 
  The attached chemical groups are indicated by small grey spheres.
  (d) Same as (c) but for $\cal{BB}$ model of iMXene functionalized with chemical groups.}
  \label{fig:imax}
\end{figure}

One of the unique structural characteristics of MAX phases is that they can be formed in various solid solutions with pure or mixtures of M, A, and X elements
~\cite{Z.Sun2003,M.F.Cover2009,M.Dahlqvist2010,Y.Mo2012,S.Aryal2014,M.Khazaei2014_1,M.Khazaei2014_2,M.Dahlqvist2015,A.Talapatra2016,M.Ashton2016_1,R.Arroyave2017}. 
In addition to traditional M$_{n+1}$AX$_{n}$ MAX phases, several sets of crystalline ordered 
double transition metal MAX phases, 
M$'_2$M$''$AX$_2$, M$'_2$M$''_2$AX$_3$~\cite{B.Anasori2015_1,R.Meshkian2017}, 
and (M$'_{2/3}$M$''_{1/3}$)$_2$AX~\cite{Q.Tao2017,M.Dahlqvist2017,R.Meshkian2018}, have been discovered recently. 
In M$'_2$M$''$AX$_2$ and M$'_2$M$''_2$AX$_3$, the layers are occupied purely with an M$'$, M$''$,
A, or X element. In (M$'_{2/3}$M$''_{1/3}$)$_2$AX, however, the double transition metals of M$'$ and M$''$ are ordered in-plane, as shown in Fig.~\ref{fig:imax}(a). 
Theoretically, another set of MAX phases with double transition metals 
(M$'_{1/2}$M$''_{1/2}$)$_{n+1}$AX$_n$ with a high ordering tendency has also been predicted, 
in which each layer is occupied purely with one type of element~\cite{A.Talapatra2016,M.Ashton2016_1}. 
In fact, the current experimental attempts to synthesize the latter MAX phase are quite close to 
becoming successful~\cite{E.N.Caspi2015}. These developments clearly demonstrate that many partially or 
perfectly ordered crystalline MXenes with unprecedented properties and applications will be synthesized 
in the near future~\cite{T.L.Tan2017}.

The in-plane ordered (M$'_{2/3}$M$''_{1/3}$)$_2$AX MAX phases have been named iMAX 
phases~\cite{A.Thore2017}. 
Experimentally, five such compounds 
have been already realized: (Mo$_{2/3}$Sc$_{1/3}$)$_2$AlC, (Mo$_{2/3}$Y$_{1/3}$)$_2$AlC, (W$_{2/3}$Sc$_{1/3}$)$_2$AlC, (W$_{2/3}$Y$_{1/3}$)$_2$AlC,
and (V$_{2/3}$Zr$_{1/3}$)$_2$AlC~\cite{Q.Tao2017,M.Dahlqvist2017,R.Meshkian2018}. 
In analogy to the 2D MXenes, the 2D layered structure derived from iMAX phases can 
accordingly be called an iMXene. 
Here, on the basis of first-principles calculations, 
we systematically investigate the electronic, mechanical, and 
dynamic properties of the experimentally synthesized iMAX phases of 
(Mo$_{2/3}$Sc$_{1/3}$)$_2$AlC, (Mo$_{2/3}$Y$_{1/3}$)$_2$AlC, (W$_{2/3}$Sc$_{1/3}$)$_2$AlC, (W$_{2/3}$Y$_{1/3}$)$_2$AlC, and (V$_{2/3}$Zr$_{1/3}$)$_2$AlC 
and their possible 2D iMXene derivatives, 
(Mo$_{2/3}$Sc$_{1/3}$)$_2$C, (Mo$_{2/3}$Y$_{1/3}$)$_2$C, (W$_{2/3}$Sc$_{1/3}$)$_2$C, (W$_{2/3}$Y$_{1/3}$)$_2$C, and (V$_{2/3}$Zr$_{1/3}$)$_2$C 
[for the crystal structure, see Fig.~\ref{fig:imax}(b)]. 
All of the considered iMAX phases 
are found to be metallic, in accordance with a previous theoretical study~\cite{A.Thore2017}. 
Single-layered iMXenes without functionalization 
are also found to be metallic. However, we find that the single-layered (Mo$_{2/3}$Sc$_{1/3}$)$_2$C, (Mo$_{2/3}$Y$_{1/3}$)$_2$C, (W$_{2/3}$Sc$_{1/3}$)$_2$C, and (W$_{2/3}$Y$_{1/3}$)$_2$C iMXenes functionalized with O 
are semiconductors with indirect band gaps. 
Moreover, we predict that these iMXenes exhibit giant piezoelectricity with a piezoelectric coefficient of 
$\sim45$$\times10^{-10}$~C/m. Therefore, the O-functionalized iMXenes are promising for exceptional piezoelectric 
applications. 


The rest of this paper is organized as follows. 
The calculation method is briefly described in Sec.~\ref{sec:method}. 
The electronic properties of the iMAX phases are reported in Sec.~\ref{sec:iMAX}, 
and the structural stability of iMXenes and  
their electronic, mechanical, and piezoelectric 
properties are examined in Sec.~\ref{sec:iMXene}. 
Finally, the paper is concluded in Sec.~\ref{sec:conclusion}.

 \section{Calculation method}  {\label{sec:method}}
    
First-principles calculations based on density-functional theory are performed 
using the VASP code~\cite{vasp1996}. The exchange-correlation energies are taken into account within the 
generalized gradient approximation (GGA) via Perdew\textemdash Burke\textemdash Ernzerhof functional~\cite{pbe1996}. 
The wave functions are constructed by the projected augmented wave method with a plane wave cutoff 
energy of 700 eV. For optimizing the atomic structure of the conventional unit cell 
of iMAX phases [see Fig.~\ref{fig:imax}(a)], 
at least 6$\times$12$\times$4 Monkhorst Pack $k$ points~\cite{monkhorst1976} are considered to 
integrate the Brillouin zone. The mechanical properties of the derived iMXenes are calculated using 
14$\times$10$\times$1 $k$ points. The energy and residual force convergence criteria are set 
to be $10^{-8}$ eV/cell and 0.0005 eV/\AA, respectively. 
In order to avoid any interaction between the single layer with its images along the $c$ axis, 
we have used a vacuum space of more than 22~\AA. 
Because the multilayered structures of iMXenes is not considered, we do not apply 
any special treatment for the van der Waals interactions.
The phonon calculations are performed 
by employing the finite displacement method. The displaced atomic configurations are obtained 
using the PHONOPY package~\cite{phonopy2008}.

The adsorption energy $E_{\rm a}$ per termination group T (T = F, O, and OH) is defined as 
$E_{\rm a} = (E_{\rm tot}-E_{\rm p}-2E_{\rm T})/2$, where 
 $E_{\rm tot}$, $E_{\rm p}$, and $E_{\rm T}$ are the total energies of the functionalized iMXene, 
 the pristine iMXene, and a termination group, respectively. 
 $E_{\rm T}$ is evaluated from the total energies of stable O$_2$, H$_2$, and F$_2$ molecules.

\begin{table}
\caption{Experimental and theoretical lattice parameters for experimentally synthesized iMAX phases 
in a base-centered monoclinic cell with $C2/c$ symmetry and thus $\alpha = \gamma = 90^{\circ}$ 
(see Fig.~\ref{fig:imax}).}
\begin{tabular}{lcccc}
\hline
\hline
iMAX phase                                                   &    $a$ (\AA)              & $b$ (\AA)             &  $c$ (\AA)          & $\beta$ ($^{\circ}$) \\
\hline
(Mo$_{2/3}$Sc$_{1/3}$)$_2$AlC                    &                                 &                              &                           &             \\
Experiment~\cite{Q.Tao2017}                                                    &            9.335             &    5.391            &  13.861             &      103.191          \\
This study                                                      &          9.349              &      5.417        &    13.925             &     103.558         \\
(Mo$_{2/3}$Y$_{1/3}$)$_2$AlC                    &                                   &                             &                            &        \\
Experiment~\cite{M.Dahlqvist2017}                                                    &             9.679            &    5.285                &  14.076                &       103.359 \\
This study                                                      &             9.569           &      5.540              &  14.126                 &        103.655      \\

 (W$_{2/3}$Sc$_{1/3}$)$_2$AlC                    &                                 &                              &                           &             \\
 Experiment~\cite{R.Meshkian2018}             &    9.368                     &   5.404              &   13.960         &    $-$              \\
 This study                                                     &     9.323              &      5.395        &    13.976          &     103.441         \\
(W$_{2/3}$Y$_{1/3}$)$_2$AlC                  &                                   &                             &                            &        \\
 Experiment~\cite{R.Meshkian2018}           &      9.510                    &    5.490               &   14.220             &    $-$    \\
 This study                                                      &       9.551                 &     5.524               &    14.136                &      103.617      \\

(V$_{2/3}$Zr$_{1/3}$)$_2$AlC                      &                   &            &              &             \\
Experiment~\cite{M.Dahlqvist2017}                                                    &            9.172         &      5.281              &    13.642              &   103.071         \\
This study                                                      &            9.160         &     5.275               &      13.617           &   103.069           \\
\hline
\hline
\end{tabular}
\label{tab:experiment}
\end{table}

In order to examine the mechanical properties of pristine and functionalized iMXenes,
the elastic strain energy per unit area is first calculated by applying in-plane strains in various directions in the 
range $-2\%\leq\varepsilon\leq2\%$ with an increment of 0.5$\%$, 
 where $\varepsilon=\frac{a-a_0}{a_0}\times100\%$ and 
$a_0$ and $a$ are the lattice parameters before and after applying the strain, respectively. 
Then, the elastic constant $C_{ij}$ is obtained by fitting the change in the total energy versus 
the applied strain with a second-order polynomial. This is carried out using the VASPKIT code~\cite{V.Wang2017}. 
Note that, following the standard Voigt notation~\cite{R.C.Andrew2012}, the elastic strain 
energy $E_{\rm s}$ per unit area can be expressed as 
$E_{\rm s}=\frac{1}{2}C_{11}\varepsilon_{xx}^2+\frac{1}{2}C_{22}\varepsilon_{yy}^2+\frac{1}{2}C_{12}\varepsilon_{xx}\varepsilon_{yy}+2C_{66}\varepsilon_{xy}^2$ for a 2D system, 
where $\varepsilon_{xx}$, $\varepsilon_{yy}$, and $\varepsilon_{xy}$ are tensile strains with $x$ and $y$ directions that 
are chosen along the $a$ and $b$ lattice vectors, respectively~\cite{S.Q.Wang2003}. 
The coefficient of the piezoelectric tensor, $e_{ij}$, is calculated on the basis of density-functional perturbation 
theory (DFPT)~\cite{R.W.Nunes2001}, as implemented in the VASP code. 
Because of the crystal symmetry studied here, $e_{11}$ and $e_{12}$
are the only independent piezoelectric coefficients. The $ij$ component of the piezoelectric strain tensor  
$d_{ij}$~\cite{R.Fei2015} is finally obtained as 
\begin{equation}
d_{11} = \frac{e_{11}C_{22}-e_{12}C_{12}}{C_{11}C_{22}-C^{2}_{12}}
\end{equation}
and
\begin{equation}
d_{12} = \frac{e_{12}C_{11}-e_{11}C_{12}}{C_{11}C_{22}-C^{2}_{12}}. 
\end{equation}

 \section{Results and discussion}  {\label{sec:result}}

\subsection{iMAX phases} {\label{sec:iMAX}

The iMAX phase possesses a base-centered monoclinic structure with $C2/c$ symmetry.
The calculated lattice parameters of five iMAX phases, 
(Mo$_{2/3}$Sc$_{1/3}$)$_2$AlC, (Mo$_{2/3}$Y$_{1/3}$)$_2$AlC, 
 (W$_{2/3}$Sc$_{1/3}$)$_2$AlC, (W$_{2/3}$Y$_{1/3}$)$_2$AlC,
and (V$_{2/3}$Zr$_{1/3}$)$_2$AlC, are summarized in Table~\ref{tab:experiment}. 
These results are in excellent agreement with the available experimental data~\cite{Q.Tao2017,M.Dahlqvist2017}. 
Note that, following the experimental reports, the lattice parameters in Table~\ref{tab:experiment} are 
for the conventional cell, which is two times larger than the primitive cell.

In order to check the dynamic stabilities of these iMAX phases, the vibrational properties are 
investigated. Figure~\ref{fig:imaxphonon} shows the calculated phonon 
dispersion and projected density of states for the conventional cell of (Mo$_{2/3}$Y$_{1/3}$)$_2$AlC 
as an example.  The phonon results for the other iMAX phases are also provided in the Supplemental Material~\cite{supplementary}.
None of the considered iMAX phases have a negative phonon frequency, indicating their structural 
stability at zero temperature. The projected phonon density of states clearly suggests 
that phonon modes related to the M$'-$C or M$''-$C vibrations appear above 12 THz and have higher 
frequencies than the phonon modes related to the M$'-$M$'$, M$'-$M$''$, M$'-$Al, M$''-$Al, or Al$-$Al 
vibrations~\cite{A.Thore2017}. This implies that, in analogy to the other members of the MAX phase family, 
the force constants of the M$'-$C or M$''-$C bonds are higher than those of the other bonds~\cite{M.Khazaei2018}. 
In other words, the M$'-$C or M$''-$C bonds are stronger than the M$'-$M$'$, M$'-$M$''$, or Al$-$Al bonds. 
If the bonding among M$'$, M$''$, and C is stronger than that with Al, there is a higher chance for 
the successful exfoliation of iMAX phases into iMXenes~\cite{M.Khazaei2018}. 
In fact, currently, (Mo$_{2/3}$Sc$_{1/3}$)$_2$AlC and (Mo$_{2/3}$Y$_{1/3}$)$_2$AlC have been 
exfoliated to Mo$_{1.33}$C experimentally~\cite{Q.Tao2017,H.Lind2017}. Likewise, 
(W$_{2/3}$Sc$_{1/3}$)$_2$AlC and (W$_{2/3}$Y$_{1/3}$)$_2$AlC have been 
exfoliated to W$_{1.33}$C experimentally~\cite{R.Meshkian2018}.

   \begin{figure}[t]
\centering
  \includegraphics[scale=0.45]{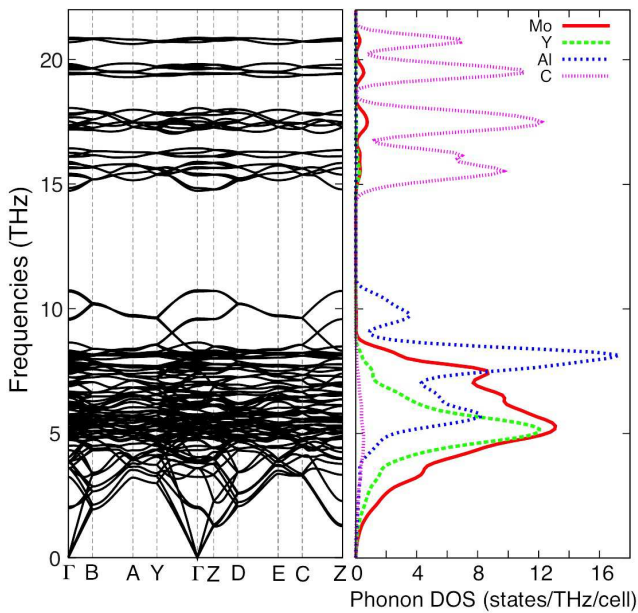}
  \caption{Phonon spectrum and phonon density of states (DOS) for (Mo$_{2/3}$Y$_{1/3}$)$_2$AlC. 
$\Gamma(0, 0, 0)$, B$(-0.5, 0, 0)$, A$(-0.5, 0.5, 0)$, Y$(0, 0.5, 0)$, Z$(0, 0, 0.5)$, D$(-0.5, 0, 0.5)$, 
E$(-0.5, 0.5, 0.5)$, and C$(0, 0.5, 0.5)$ are high symmetric points of the Brillouin zone.
}
  \label{fig:imaxphonon}
\end{figure}

 \begin{figure*}[t]
\centering
  \includegraphics[scale=0.4]{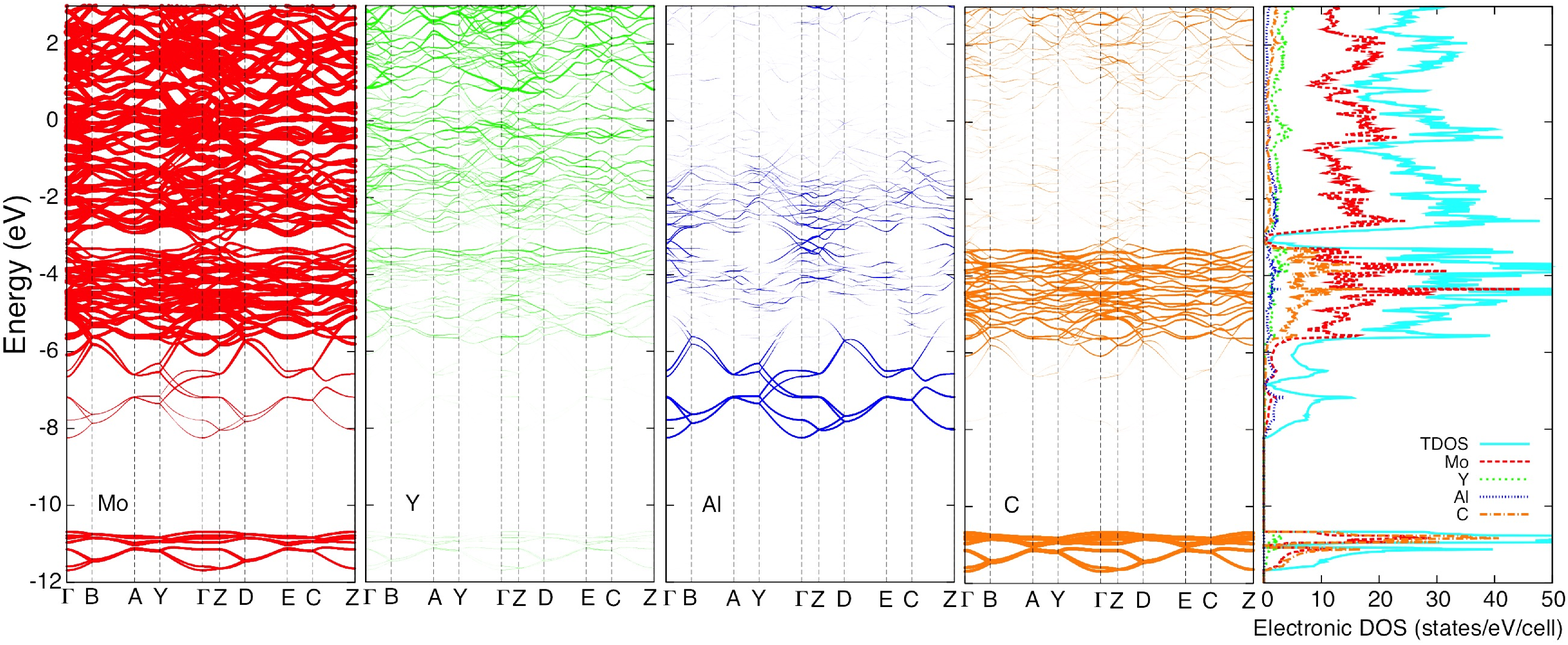}
  \caption{Band structure and density of states (DOS) of (Mo$_{2/3}$Y$_{1/3}$)$_2$AlC, projected onto the 
  compositional elements, Mo, Y, Al, and C. 
  TDOS stands for total density of states. The Fermi energy is set at zero.}
  \label{fig:imaxelectronic}
\end{figure*}
 
We also examine the electronic structures of (Mo$_{2/3}$Sc$_{1/3}$)$_2$AlC, (Mo$_{2/3}$Y$_{1/3}$)$_2$AlC, 
(W$_{2/3}$Sc$_{1/3}$)$_2$AlC, (W$_{2/3}$Y$_{1/3}$)$_2$AlC,
and (V$_{2/3}$Zr$_{1/3}$)$_2$AlC. 
For instance, the results for (Mo$_{2/3}$Y$_{1/3}$)$_2$AlC are shown in Fig.~\ref{fig:imaxelectronic}. 
The band structures and density of states clearly show that, similar to the other MAX phases, the iMAX phases are metallic. 
This can be simply explained by considering that in MAX phases, including iMAX phases, the transition metal 
atoms are connected in each layer, resulting in the dominant contribution of the $d$ orbitals of the transition metals 
to the states near the Fermi energy and consequently to the electrical conductivity of these phases. 
The projected DOSs of the M$'$ and M$''$ transition metals are similarly distributed
around the Fermi energy. However, the M$'$ atoms have larger
contribution since, in (M$'_{2/3}$M$''_{1/3}$)$_2$AX, the number of
M$'$ atoms is larger than that of M$''$ atoms.

The projected density of states (PDOS) in Fig.~\ref{fig:imaxelectronic} also shows that there is significant 
hybridization between the C atoms and the transition metals M$'$ and M$''$. 
The states between $-6$ and $-3$~eV ($-12$ and $-10$~eV) result from the hybridization 
between the $p$ ($s$) orbitals of 
C and the $d$ orbitals of M$'$ and M$''$. There also exists hybridization between the $p$ orbitals of Al and the $d$ 
orbitals of M$'$ and M$''$, appearing in a broad range of energies between $-6$ and $-1$~eV. 
However, this hybridization is not as strong as the hybridization between 
the atomic orbitals of C and M$'$ or M$''$.

In both the MAX and iMAX phases, the energy gaps at around $-9$ eV separate
the valence bands from other low lying states such as the M or C $s$ bands. Just
above these energy gaps, the states at the lowest part of the valence bands
result from the hybridization between the atomic orbitals of Al atoms in
the Al$-$Al bonds (see the Supplemental Material~\cite{supplementary}). In typical MAX
phases where the Al atoms are located on a uniform hexagonal lattice, the
energy bands in this energy region show a parabolic dispersion, and the total
DOSs do not show notable features. In the iMAX phases, on the other hand,
additional features appear in the energy bands and DOSs. For example, the
total DOS becomes almost zero around $-7$ eV for (Mo$_{2/3}$Y$_{1/3}$)AlC (see Fig.~\ref{fig:imaxelectronic}). 
Such a pseudogap originates from the Kagome-like alignment of Al
atoms. Thus, these features can be considered as a kind of fingerprint
of the iMAX phases. In addition, the Al$-$Al bond lengths in the 
(M$'_{2/3}$M$''_{1/3}$)$_2$AC iMAX phases are slightly shorter than those in 
M$'_2$AC or M$''_2$AC; therefore, we expect that the Al$-$Al bonds in the iMAX
phases should be slightly stronger than those in other MAX phases.

Recently, we have performed a very intensive study on the possibility of exfoliation of 82 
different MAX-phase compounds into 2D MXenes~\cite{M.Khazaei2018}. According to our analysis, 
MAX phases with exfoliation energies less than 0.205 eV/\AA$^2$ have a very high chance to be exfoliated into 2D MXenes. 
We have performed a similar analysis of the exfoliation 
energies of iMAX phases into 2D iMXenes. The exfoliation energies of (Mo$_{2/3}$Sc$_{1/3}$)$_2$AlC, (Mo$_{2/3}$Y$_{1/3}$)$_2$AlC, 
(W$_{2/3}$Sc$_{1/3}$)$_2$AlC, (W$_{2/3}$Y$_{1/3}$)$_2$AlC,
and (V$_{2/3}$Zr$_{1/3}$)$_2$AlC to their corresponding iMXenes, (Mo$_{2/3}$Sc$_{1/3}$)$_2$C, (Mo$_{2/3}$Y$_{1/3}$)$_2$C, 
(W$_{2/3}$Sc$_{1/3}$)$_2$C, (W$_{2/3}$Y$_{1/3}$)$_2$C, and 
(V$_{2/3}$Zr$_{1/3}$)$_2$C, are estimated to be around 0.156, 0.172, 0.177, 0.162, and 0.185 eV/\AA$^2$, respectively. 
Since all obtained exfoliation energies are less than 0.205 eV/\AA$^2$, we predict, according to our 
previous analysis~\cite{M.Khazaei2018}, that the above iMAX 
phases can be exfoliated into 2D iMXenes, 
as indeed observed in experiments~\cite{Q.Tao2017,R.Meshkian2018,H.Lind2017}.

\subsection{iMXenes} \label{sec:iMXene}

Upon chemical exfoliation, the MAX phases can be exfoliated into single- or multilayered MXenes, and their surfaces 
are terminated with mixture of F, OH, and O groups~\cite{M.Naguib2011,M.Naguib2012,I.Persson2017} 
owing to the reactivity of the surface transition metals. 
Theoretical studies indicate that fully saturated MXenes are thermodynamically more favorable than 
partially saturated ones~\cite{M.Khazaei2013,M.Ashton_2}. 
Thermodynamic simulations show that, by proper control of the chemical potential, the surfaces of MXenes can be 
saturated with a particular type of chemical group~\cite{M.Khazaei2013,M.Ashton_2}. 
Although the synthesis of MXenes with a particular surface termination is still a challenge in 
wet chemistry experiments, there have been developments regarding the control of the surface termination of 
MXenes by using a thermal process~\cite{I.Persson2017}. 
Using this technique, it might be possible to synthesize pristine MXenes. 
Therefore, we theoretically study the structural stability of iMXenes and their electronic properties.

\subsubsection{Atomic structure and dynamic stability}

It has been repeatedly shown that exfoliated MXenes possess many interesting properties and 
applications that their original MAX phases do not show~\cite{B.Anasori2017_1,M.Khazaei2017}. 
For instance, although all MAX phases are metallic, upon exfoliation and proper surface functionalization, 
some of them turn into semiconductors~\cite{M.Khazaei2013} or 
topological insulators~\cite{H.Weng2015,M.Khazaei2016_2,Y. Liang2017,L.Li2016,C.Si2016,C.Si2016_2}. 
Such interesting properties and applications motivate us to investigate the electronic properties of iMXenes 
in detail. 
 
 \begin{figure*}[t]
\centering
  \includegraphics[scale=0.45]{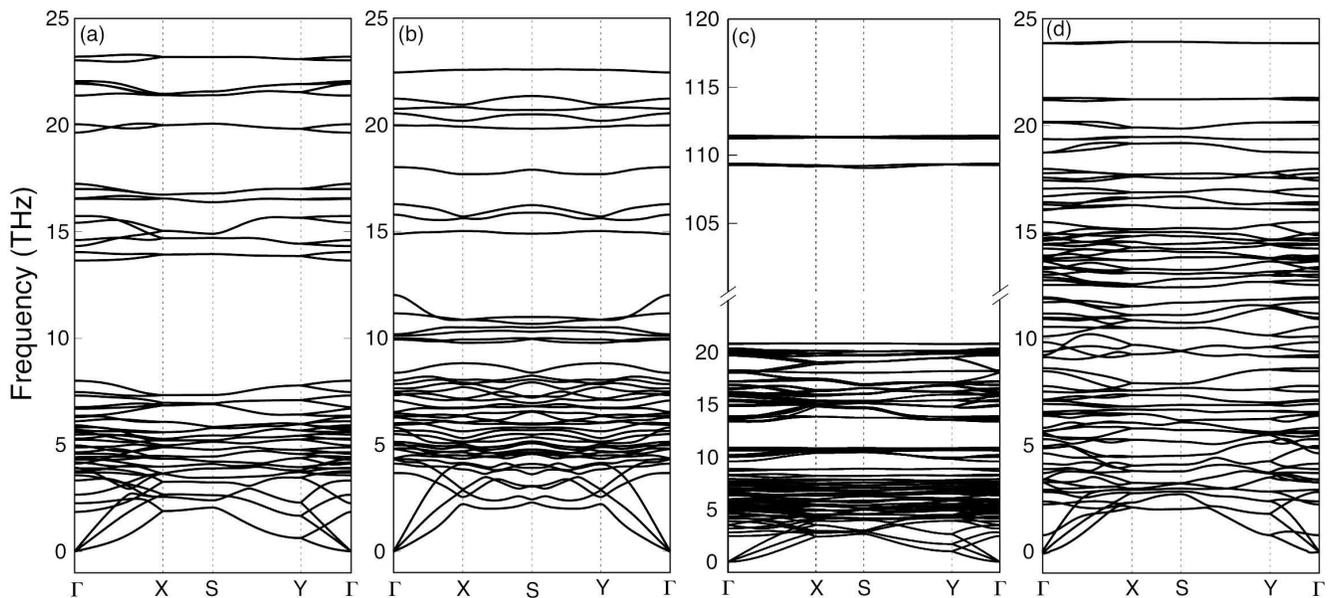}
  \caption{
  Phonon band structures of (a) pristine (Mo$_{2/3}$Y$_{1/3}$)$_2$C, (b) (Mo$_{2/3}$Y$_{1/3}$)$_2$CF$_2$, 
  (c) (Mo$_{2/3}$Y$_{1/3}$)$_2$C(OH)$_2$, and (d) (Mo$_{2/3}$Y$_{1/3}$)$_2$CO$_2$ iMXenes. 
$\Gamma(0, 0, 0)$, X$(0.5, 0, 0)$, S$(0.5, 0.5, 0)$, and Y$(0, 0.5, 0)$ are high symmetric points of the Brillouin zone.
{\color{red}} 
}
  \label{fig:phono-imxene}
\end{figure*}

The electronic structures of MXenes are significantly affected by the type and position of the chemical groups 
absorbed on the surfaces of MXenes~\cite{M.Khazaei2013}. 
Hence, in order to propose the lowest energy atomic configuration for functionalized iMXenes, 
it is necessary to investigate the stability of various iMXenes with functionalization at different adsorption sites. 
Although the atomic configurations of the Al atoms in typical MAX phases (with Al atoms forming a hexagonal lattice) 
and in the iMAX phases (with Al atoms forming a Kagome-like lattice) are different, the pristine iMXene 
(M$'_{2/3}$M$''_{1/3}$)$_2$C after removing Al atoms possesses a surface structure similar to 
that of the pristine M$_2$C MXene with two types of hollow sites, $\cal{A}$ and $\cal{B}$, as indicated 
in Fig.~\ref{fig:imax}(b).

For hollow site $\cal{A}$ ($\cal{B}$), there is no (there is a) C atom underneath it. 
Previous calculations have revealed that when the surfaces of MXenes are saturated with F, OH, or O, 
these chemical groups prefer to be adsorbed at the hollow sites rather than adsorbed directly on top of 
the transition metals~\cite{M.Khazaei2013,M.Khazaei2014_3}. 
Therefore, depending on the positions of the adsorbed atoms at the hollow sites, there are three possibilities for 
the functionalized iMXenes, i.e., $\cal{AA}$, $\cal{BB}$, and $\cal{AB}$ model structures. 
In model structure $\cal{AA}$, the opposite sides of surfaces are saturated with chemical groups adsorbed at 
$\cal{A}$-type hollow sites [see Fig.~\ref{fig:imax}(c)]. 
In model structure $\cal{BB}$, the opposite sides of surfaces are saturated with chemical groups adsorbed 
at $\cal{B}$-type hollow sites [see Fig.~\ref{fig:imax}(d)]. 
In model structure $\cal{AB}$, one side of the surface is saturated with chemical groups adsorbed at 
$\cal{A}$-type hollow sites, and the opposite side is saturated with chemical groups adsorbed 
at $\cal{B}$-type hollow sites. 
Note that each unit formula of (M$'_{2/3}$M$''_{1/3}$)$_2$C requires two termination groups (T = F, OH, and O) 
for the full surface termination in (M$'_{2/3}$M$''_{1/3}$)$_2$CT$_2$. 

Table~\ref{tab:stability} summarizes the calculated total energies of the optimized lattice structures 
of various iMXenes for the three different model structures 
functionalized with F, OH, and O groups. 
We find that for all iMXenes studied here, regardless of the type of functionalization, model structure $\cal{AA}$ has 
the lowest energy. Accordingly, we also show the adsorption energy $E_{\rm a}$ 
of the functional groups on the iMXenes with the most stable model structure $\cal{AA}$ in Table~\ref{tab:stability}. 
These adsorption energies are large negative values, indicating the formation of strong chemical bonds 
between the surface and the functional groups~\cite{M.Khazaei2013}. 
In fact, the functional groups strengthen the stability of MXenes by saturating the nonbonding 
valence electrons of the transition metals~\cite{T.Hu2017}.

\begin{table}
\caption{
Total energy differences (eV per functional group) of different model structures 
$\cal{AA}$, $\cal{AB}$, and $\cal{BB}$ for 
(M$'_{2/3}$M$''_{1/3}$)$_2$CT$_2$ with T= F, OH, and O. 
The total energy of the model structure with the lowest energy is set to be zero. 
$E_{\rm a}$ is the adsorption energy (eV per functional group).}
\begin{tabular}{lcccc}
\hline
\hline
Functionalized iMXene                                                             &    $\cal{AA}$             &  $\cal{AB}$                 &  $\cal{BB}$              &  $E_{\rm a}$ \\
\hline
(Mo$_{2/3}$Sc$_{1/3}$)$_2$CF$_2$                     &             0.000                    &       0.160                      &     0.030       & $-4.729$      \\
(Mo$_{2/3}$Y$_{1/3}$)$_2$CF$_2$                      &                 0.000             &              0.105                &      0.191            &  $-4.689$     \\
 (W$_{2/3}$Sc$_{1/3}$)$_2$CF$_2$                    &            0.000                    &    0.014   &    0.189          &    $-4.561$    \\
(W$_{2/3}$Y$_{1/3}$)$_2$CF$_2$                     &               0.000  &      0.062     &     0.061           &    $-4.530$   \\
(V$_{2/3}$Zr$_{1/3}$)$_2$CF$_2$                        &                0.000               &         0.014                     &    0.318             & $-4.982$  \\

(Mo$_{2/3}$Sc$_{1/3}$)$_2$CO$_2$                    &               0.000           &          0.154                &     0.307             &  $-4.173$  \\
(Mo$_{2/3}$Y$_{1/3}$)$_2$CO$_2$                     &                  0.000               &            0.053                  &      0.221          &  $-4.054$ \\
(W$_{2/3}$Sc$_{1/3}$)$_2$CO$_2$                    &           0.000                     &     0.168                       &       0.370     &   $-4.533$   \\
(W$_{2/3}$Y$_{1/3}$)$_2$CO$_2$                     &                0.000            &       0.142                  &        0.324      &  $-4.433$     \\                                                          
(V$_{2/3}$Zr$_{1/3}$)$_2$CO$_2$                       &                  0.000            &           0.193                 &       0.430          & $-4.982$ \\   
  
(Mo$_{2/3}$Sc$_{1/3}$)$_2$C(OH)$_2$              &             0.000          &         0.1073                    &      0.294             & $-4.166$ \\
(Mo$_{2/3}$Y$_{1/3}$)$_2$C(OH)$_2$                &                 0.000     &              0.048                &          0.332         & $-4.074$\\
(W$_{2/3}$Sc$_{1/3}$)$_2$C(OH)$_2$                     &             0.000                     &     0.086                     &    0.270   &   $-4.082$     \\
(W$_{2/3}$Y$_{1/3}$)$_2$C(OH)$_2$                    &                0.000             &         0.016                &      0.344        &  $-3.988$    \\
(V$_{2/3}$Zr$_{1/3}$)$_2$C(OH)$_2$                  &                 0.000              &            0.169                  &      0.219     &   $-4.363$     \\                                                                     
\hline
\hline
\end{tabular}
\label{tab:stability}
\end{table}

We also investigate the dynamic stability of the predicted iMXene structures on the basis of a set of phonon 
calculations. The results for pristine and functionalized 
(Mo$_{2/3}$Y$_{1/3}$)$_2$C are shown in Fig.~\ref{fig:phono-imxene} as an example. 
 The phonon results for other iMXenes are provided in the Supplemental Material~\cite{supplementary}.
The phonon frequencies of most of the studied iMXenes are positive, indicating that their structures are dynamically stable.
Few of them have small negative phonon frequencies, which are generally observed in the phonon spectra of 2D systems 
and can possibly be removed by applying a small strain or deposition onto a proper  substrate~\cite{Y.Li2015,E.S.Penev2016,W.Sun2016,J.Lei2017,J.He2017,M.Mushtaq2017}. 
Such small negative phonon frequencies could also be an artifact of the numerical inaccuracy due to the limited 
supercell size, cutoff energy, or number of $k$ points, or reflect the actual 
lattice instability towards large wave undulations of 2D materials. 
However, (V$_{2/3}$Zr$_{1/3}$)$_2$CO$_2$ shows clear negative 
phonon modes (see the Supplemental Material~\cite{supplementary}), 
which indicates the model structure $\cal{AA}$ of this iMXexe is not dynamically stable. 
Therefore, the second lowest energy structure of this iMXexe, i.e., model structure $\cal{AB}$, is the stable structure 
for (V$_{2/3}$Zr$_{1/3}$)$_2$CO$_2$, which shows all positive phonon spectra.

 \begin{figure*}[t]
\centering
 \includegraphics[scale=0.45]{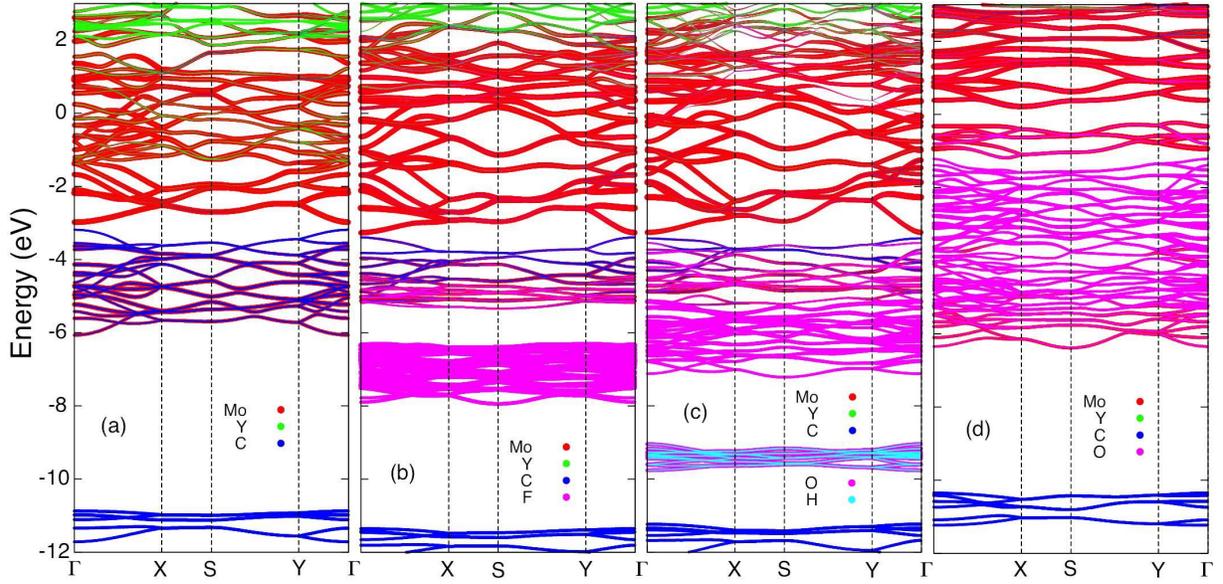}
  \caption{Projected band structures for (a) pristine (Mo$_{2/3}$Y$_{1/3}$)$_2$C, 
  (b) (Mo$_{2/3}$Y$_{1/3}$)$_2$CF$_2$, (c) (Mo$_{2/3}$Y$_{1/3}$)$_2$C(OH)$_2$, and 
  (d) (Mo$_{2/3}$Y$_{1/3}$)$_2$CO$_2$ iMXenes. 
$\Gamma(0, 0, 0)$, X$(0.5, 0, 0)$, S$(0.5, 0.5, 0)$, and Y$(0, 0.5, 0)$
 are high symmetric points of the Brillouin zone. The Fermi energy is set at zero. 
}
  \label{fig:imxene-dos}
\end{figure*}

\subsubsection{Electronic structures}

As an example of the electronic structures of iMXenes, Fig.~\ref{fig:imxene-dos} shows the results for pristine 
and functionalized (Mo$_{2/3}$Y$_{1/3}$)$_2$C with F, OH, and O. 
We find that the pristine (Mo$_{2/3}$Y$_{1/3}$)$_2$C iMXexe is metallic. 
It remains metallic upon functionalization with F and OH, but it becomes semiconducting when it is 
functionalized with O. In more detail,
as seen in Fig.~\ref{fig:imxene-dos}(a), pristine (Mo$_{2/3}$Sc$_{1/3}$)$_2$C is metallic, 
and the states between $-3$ and 3 eV are dominated mainly by the $d$ orbitals of Mo atoms. 
These states are formed owing to the hybridization among the $d$ orbitals of Mo atoms. 
The states between $-6$ and $-3$ eV result from 
the hybridization between the $p$ orbitals of C atoms and the $d$ orbitals of the transition metals. 
The $s$ orbitals of C atoms are hybridized with the $d$ orbitals of the transition metals in a much deeper energy 
region between $-12$ and $-10$ eV.

The functionalization of pristine iMXenes with F, OH, or O induces hybridization between the atomic orbitals 
of the attached functional groups and the transition metals, and the new hybridized states appear between 
the $p$ and $s$ subbands of the C atoms. Owing to the formation of these hybridized states and the simultaneous 
electron transfer from the transition metals to F or O atoms,
the Fermi energy shifts downwards to lower energies upon functionalization (see Fig.~\ref{fig:imxene-dos}).
The Fermi energy shift for the O-terminated iMXenes is larger than that for the F-terminated ones simply
because of the different valencies of these two atoms.

Among the pristine and functionalized iMXenes studied here, only (Mo$_{2/3}$Y$_{1/3}$)$_2$CO$_2$, 
(Mo$_{2/3}$Sc$_{1/3}$)$_2$CO$_2$, (W$_{2/3}$Y$_{1/3}$)$_2$CO$_2$, and  (W$_{2/3}$Sc$_{1/3}$)$_2$CO$_2$ are semiconductors 
with indirect band gaps of 0.45, 0.04, 0.625, and 0.675 eV, respectively. The origin of these band gaps may be attributed 
to the splitting of Mo $d$-orbital-dominated bands.
In functionalized MXenes, the transition metals are surrounded with C atoms and the attached chemical 
groups (F, OH, or O).
Such a local environment splits the $d$ bands into subbands~\cite{T.Hu2017}. 
More precisely,
the states near the Fermi energy can be assigned as t$_{2g}$ bands hybridized with C, F, or O $p$ orbitals 
via $dp\pi$ coupling. 
In the case of  (Mo$_{2/3}$Y$_{1/3}$)$_2$CO$_2$,  (Mo$_{2/3}$Sc$_{1/3}$)$_2$CO$_2$, (W$_{2/3}$Y$_{1/3}$)$_2$CO$_2$, and  (W$_{2/3}$Sc$_{1/3}$)$_2$CO$_2$,
the t$_{2g}$ 
bands are further split by the low symmetry structure, resulting in finite band gaps. 
It is known that the DFT method with the GGA-PBE functional underestimates the band gaps, while hybrid functionals, \textit{e.g.}, 
Heyd$-$Scuseria$-$Ernzerhof (HSE06)~\cite{J.Heyd2003,J.Heyd2006},
might give a better estimate. Hence, we also examine the band gaps by using the HSE06 functional (see the Supplemental Material~\cite{supplementary}). 
As expected, in the HSE06 level of theory, (Mo$_{2/3}$Sc$_{1/3}$)$_2$CO$_2$, 
(Mo$_{2/3}$Y$_{1/3}$)$_2$CO$_2$, (W$_{2/3}$Sc$_{1/3}$)$_2$CO$_2$ and 
(W$_{2/3}$Y$_{1/3}$)$_2$CO$_2$ obtain larger band gaps of 0.58, 1.23, 1.3, and 1.3 eV,  respectively, than those obtained from the GGA-PBE method.
 

\subsubsection{Mechanical properties}

\begin{table*}
\small
  \caption{Calculated elastic constants $C_{ij}$ (N/m), Young's modulus $Y_\alpha$ (N/m) 
  along the $\alpha$ ($=x$ and $y$) direction, and 
  Poisson's ratio $\nu_{\alpha\beta}$, the strain response in the $\alpha$ direction to the strain applied along 
  the $\beta$ direction, for various iMXenes.
}
  \label{tab:mechanical}
  \begin{tabular*}{\textwidth}{@{\extracolsep{\fill}}lccccccccc}
\hline
\hline
 iMXene                                                              &   $C_{11}$  &    $C_{22}$  &   $C_{12}$   &   $C_{66}$ &  $Y_x$  &   $Y_y$   &   $\nu_{xy}$   &   $\nu_{yx}$  &  Space group    \\
\hline
(Mo$_{2/3}$Sc$_{1/3}$)$_2$C                           &       130.944  &  121.737   &  22.220        & 39.465       & 126.888  &  117.966   &  0.183     &   0.170     &  $C2/m$ (No. 12)  \\
(Mo$_{2/3}$Y$_{1/3}$)$_2$C                              &  114.102        &  74.386     & -2.955         & 42.635       & 113.984  &  74.309     & -0.040     &    -0.026   &  $C2/m$ (No. 12)  \\                  
 (W$_{2/3}$Sc$_{1/3}$)$_{2}$C       &  111.633 & 115.741 & 18.825 & 52.012 & 108.571 & 112.566 &  0.163 & 0.169  &     $C2/m$ (No. 12)        \\  
(W$_{2/3}$Y$_{1/3}$)$_{2}$C       &  89.747 & 113.441 &  36.990 & 45.595  & 77.685 & 98.195 &  0.326 & 0.412  &    $C2/m$ (No. 12)         \\ 
(V$_{2/3}$Zr$_{1/3}$)$_2$C                                &    120.676     &  91.458      &  32.128       &  43.205      & 109.390 &  82.905      &   0.351     &  0.266      &   $C2/m$ (No. 12)  \\

(Mo$_{2/3}$Sc$_{1/3}$)$_2$CF$_2$                    & 154.389 &158.156 & 59.944 & 45.577 &131.669 &134.882 &  0.379 & 0.388  & $C2/m$ (No. 12) \\
(Mo$_{2/3}$Y$_{1/3}$)$_2$CF$_2$             &136.214 & 147.712 & 54.653 & 47.224 & 115.992 & 125.783 & 0.370 & 0.401 & $C2/m$ (No. 12)  \\   
 (W$_{2/3}$Sc$_{1/3}$)$_{2}$CF$_{2}$        & 142.892 & 149.628 & 50.781 & 39.587  & 125.657 & 131.581 &  0.339 & 0.339  &    $C2/m$ (No. 12)         \\  
 (W$_{2/3}$Y$_{1/3}$)$_{2}$CF$_{2}$      & 156.307 & 155.771 &  39.698 & 45.547  & 146.189 & 145.689 &  0.255 &  0.254  &   $C2/m$ (No. 12)         \\        
(V$_{2/3}$Zr$_{1/3}$)$_2$CF$_2$               &159.566 & 159.529 & 41.999 & 58.976 & 148.509 & 148.475 & 0.263 & 0.263  &  $C2/m$ (No. 12)\\

(Mo$_{2/3}$Sc$_{1/3}$)$_2$CO$_2$  & 257.927 & 198.383 & 92.341   & 76.824  & 214.946 & 165.324 & 0.465 & 0.358  &   $C2$ (No. 5)\\         
(Mo$_{2/3}$Y$_{1/3}$)$_2$CO$_2$    & 224.802 &142.086 & 73.584 & 56.601 & 186.694 & 117.999 &  0.518 & 0.327  &  $C2$ (No. 5) \\   
 (W$_{2/3}$Sc$_{1/3}$)$_{2}$CO$_{2}$         &  279.897 & 219.678 &  90.389 & 85.458  & 242.705 & 190.488 &   0.411 &  0.323  &    $C2$ (No. 5)        \\ 
 (W$_{2/3}$Y$_{1/3}$)$_{2}$CO$_{2}$      &  250.670 & 173.314 & 79.150 & 68.260  & 214.524 & 148.323 &  0.457 & 0.316  &    $C2$ (No. 5)          \\                                                          
(V$_{2/3}$Zr$_{1/3}$)$_2$CO$_2$      & 243.120 & 223.845 & 92.442 & 66.270  & 204.944 & 188.695 & 0.413 & 0.380  &  $Cm$ (No. 8) \\             
  
(Mo$_{2/3}$Sc$_{1/3}$)$_2$C(OH)$_2$    & 164.041 & 164.822 & 49.382 & 54.335 & 149.246 & 149.957 & 0.300 & 0.301 &  $C2/m$ (No. 12)\\      
(Mo$_{2/3}$Y$_{1/3}$)$_2$C(OH)$_2$       & 141.653 & 148.998 & 46.117 & 54.075 & 127.380 & 133.984 & 0.310 & 0.326 &  $C2/m$ (No. 12)\\        
 (W$_{2/3}$Sc$_{1/3}$)$_{2}$C(OH)$_{2}$       & 155.277 &  153.032 & 48.042 & 54.723  & 140.195 & 138.169 &   0.314 &  0.309  &     $C2/m$ (No. 12)        \\ 
 (W$_{2/}$Y$_{1/3}$)$_{2}$C(OH)$_{2}$      & 143.140 & 145.001 &  36.714 & 55.714  & 133.844 & 135.584 &   0.253 & 0.256  &    $C2/m$ (No. 12)        \\    
(V$_{2/3}$Zr$_{1/3}$)$_2$C(OH)$_2$        & 170.367 & 159.457 & 40.705 & 63.418  & 159.976 & 149.732 &  0.255 & 0.239  &   $C2/m$ (No. 12)      \\          
                                                                
\hline
\hline
  \end{tabular*}
\end{table*}

As benchmark tests for elastic property calculations, we have first considered graphene and 
borophene sheets~\cite{V.Wang2017} and found that the results obtained 
for $C_{11}$, $C_{22}$, $C_{12}$, and $C_{66}$ are in excellent agreement with those reported 
previously~\cite{S.Q.Wang2003,A.J.Mannix2015}. 
Encouraged by these results, we now calculate the elastic constants of pristine and 
functionalized (Mo$_{2/3}$Sc$_{1/3}$)$_2$C, (Mo$_{2/3}$Y$_{1/3}$)$_2$C, 
(V$_{2/3}$Zr$_{1/3}$)$_2$C, (W$_{2/3}$Sc$_{1/3}$)$_{2}$C, and (W$_{2/3}$Y$_{1/3}$)$_{2}$C iMXenes. The results are summarized in Table \ref{tab:mechanical}. 
The obtained elastic constants for all of these iMXenes satisfy the Born criteria, 
$C_{11}+C_{22}-C_{12}^2>0$ and $C_{66}>0$~\cite{M.Born1954}. This implies that all of these iMXenes 
are  mechanically stable. 
Moreover, we find that all of the elastic constants of the pristine iMXenes increase with F, O, or OH functionalization, 
suggesting that the functionalized iMXenes have better mechanical properties 
than the pristine ones. 
This is in good accordance with previous studies of other pristine and functionalized 
MXenes~\cite{Z.H.Fu2016,Z.Guo2015}.

In addition, we calculate the in-plane Young moduli 
$Y_x=\frac{1}{C_{22}} (C_{11}C_{22}-C_{12}C_{21})$ and $Y_y=\frac{1}{C_{11}}(C_{11}C_{22}-C_{12}C_{21})$, 
and the Poisson's ratios $\nu_{xy}=\frac {C_{21}}{C_{22}}$ and $\nu_{yx}=\frac {C_{12}}{C_{11}}$ of 
the iMXenes~\cite{A.J.Mannix2015,Q.Wei2014}, which are summarized in Table~\ref{tab:mechanical}. 
We find that the O-functionalized iMXenes possess the highest Young's moduli, as large as almost 200 N/m, 
followed by the OH-functionalized, the F functionalized, and pristine iMXenes. 
For comparison, we also calculate the Young's moduli of graphene (340 N/m) 
and BN (279.2 N/m), which are in excellent agreement with previous results ~\cite{C.Lee2008,Q.Peng2012}. 
This suggests that iMXenes, especially the pristine ones, belong to a class of soft 2D materials.

Interestingly, the O-functionalized iMXenes possess large Poisson's ratios, in particular, 
(Mo$_{2/3}$Y$_{1/3}$)$_2$CO$_2$ whose Poisson's ratio can reach up to 0.57. 
Note that most 2D materials have a Poisson's ratio less than 0.3, e.g., 
0.18 for graphene~\cite{F.Lu2007} and 0.22 for both BN~\cite{Q.Peng2012} and MoS$_2$~\cite{J.L.Feldman1976}. 
It is also noted that the Poisson's ratio of an aluminium oxide monolayer is close to 0.68~\cite{T.T.Song2016}. 
Moreover, among the studied iMXenes, pristine (Mo$_{2/3}$Y$_{1/3}$)$_2$C shows a small negative 
Poisson's ratio, which has only been found in a few sets of 
2D materials~\cite{Z.Gao2017}. 
Although pristine MXenes and iMXenes are difficult to realize experimentally using wet chemical 
exfoliation because of their strong tendency to absob 
F, O, or OH groups, this unique property motivates experimentalists to synthesize pristine MXenes with 
other experimental techniques such as chemical vapor deposition.

\subsubsection{Piezoelectric properties}

Considering the symmetry of the iMXenes, indicated in Table~\ref{tab:mechanical}, the pristine, F-, and OH- functionalized iMXenes with $C2/m$ symmetry have centrosymmetry except O-functionalized ones with $C2$ symmetry. 
Therefore, it can expected 
the semiconducting iMXenes, (Mo$_{2/3}$Y$_{1/3}$)$_2$CO$_2$, (Mo$_{2/3}$Sc$_{1/3}$)$_2$CO$_2$, 
(W$_{2/3}$Sc$_{1/3}$)$_{2}$CO$_{2}$, and (W$_{2/3}$Y$_{1/3}$)$_{2}$CO$_{2}$
to have piezoelectric properties.
Therefore, the two independent piezoelectric stress coefficients, 
$e_{11}$ and $e_{12}$, represent the piezoelectric response under the uniaxial strains 
$\varepsilon_{xx}$ and $\varepsilon_{yy}$. 
Table~\ref{tab:piezo} summarizes the calculated clamped- and relaxed-ion piezoelectric stress coefficients~\cite{M.M.Alyoruk2015}. 
The clamped ion stands for the calculations in which the positions of ions are 
not optimized after applying a uniaxial strain.
The relaxed and clamped piezoelectric strain coefficients of $d_{11}$ and $d_{12}$, also listed in Table~\ref{tab:piezo}}, 
reflect the amount of induced polarization 
upon applying an external force and therefore represent the mechanical-electrical energy converting ratio. 
We find that the relaxed $d_{11}$ and $d_{12}$ values are notably larger for these four iMXenes than those 
predicted for transition metal dichalcogenides (estimated as 2.12$-$13.5 pmV$^{-1}$)~\cite{K.A.N.Duerloo2012,M.M.Alyoruk2015,M.N.Blonsky2015}, 
most monolayered 
materials such as BN, CaS, GaSe, GaAs, and AlSb (estimated as 0.5$-$3.0 pmV$^{-1}$)~\cite{R.Fei2015,W.Li2015,H.Zheng2015, C.Sevik2016}, 
and bulk materials including $\alpha$-quartz, wurtzite GaN,
and wurtzite AlN (estimated as 2.3, 3.1, and 5.1 pmV$^{-1}$)~\cite{R.Bechmann1958, C.M.Lueng2000}, which are widely used in industry.

In order to check the results above, we also calculate $e_{11}$ and $e_{12}$ from 
the relation $e_{ij}=\partial P_i$/$\partial\varepsilon_{j}$, where $P_i$ is the induced polarization 
evaluated by the Berry phase method along the $i$ direction in the presence of the strain 
$\varepsilon_{j}$ applied along the $j$ direction. 
The piezoelectric coefficients are obtained by a linear fitting of the change in $P_i$ versus 
$\varepsilon_{j}$, 
in which $\varepsilon_{j}$ varies between -0.01 and +0.01 in increments of 0.002.  
We find that the obtained relaxed-ion $e_{11}$ piezoelectric coefficients of
(Mo$_{2/3}$Y$_{1/3}$)$_2$CO$_2$, (Mo$_{2/3}$Sc$_{1/3}$)$_2$CO$_2$, (W$_{2/3}$Sc$_{1/3}$)$_{2}$CO$_{2}$, 
and (W$_{2/3}$Y$_{1/3}$)$_{2}$CO$_{2}$, 39.54, 43.58, 38.02, and 33.89 $\times10^{-10}$~C/m, respectively, 
are in excellent agreement with the DFPT results listed in Table~\ref{tab:piezo}. 
Therefore, these results clearly demonstrate that iMXenes have peculiar piezoelectric 
properties with $d_{11}$ one order of magnitude larger than those of commercially utilized bulk materials 
and most semiconducting 2D materials. 
Finally, we note that we have also calculated the relaxed-ion piezoelectric coefficients of Sc$_2$CO$_2$, 
the only system without centrosymmetry among the M$_2$CO$_2$ (M = Sc, Ti, Zr, and Hf) MXenes 
and found that 
$e_{11} = 3.333\times10^{-10}$ ${\rm Cm}^{-1}$ and $d_{11} = 4.137~{\rm pVm}^{-1}$. 
Therefore, we predict that the piezoelectric properties of Sc$_2$CO$_2$ are similar to those of well-known transition metal dichalcogenides, 
MoS$_2$ and MoSe$_2$~\cite{M.N.Blonsky2015},
but smaller than those of the iMXenes investigated in this study.

\begin{table}
\small
\caption{
Calculated clamped- and relaxed-ion piezoelectric stress ($e_{11}$ and $e_{12}$ in unit of 
$10^{-10}$ Cm$^{-1}$) and piezoelectric strain ($d_{11}$ and $d_{12}$ in unit of pVm$^{-1}$) 
coefficients for (Mo$_{2/3}$Sc$_{1/3}$)$_2$CO$_2$, (Mo$_{2/3}$Y$_{1/3}$)$_2$CO$_2$, 
(W$_{2/3}$Sc$_{1/3}$)$_{2}$CO$_{2}$, and (W$_{2/3}$Y$_{1/3}$)$_{2}$CO$_{2}$ iMXenes.}
\label{tab:piezo}
\begin{tabular}{lcccc}
\hline
\hline
&\multicolumn{2}{c}{Clamped-ion}&\multicolumn{2}{c}{Relaxed-ion}\\
 iMXene &   $e_{11}$  &   $e_{12}$  &   $e_{11}$   &   $e_{12}$  \\
\hline
(Mo$_{2/3}$Sc$_{1/3}$)$_2$CO$_2$  & 11.76 & $-4.80$ & 44.55 & $-10.99$\\         
(Mo$_{2/3}$Y$_{1/3}$)$_2$CO$_2$    & 30.52 & $-1.64$ & 40.33 & $-4.89$\\
(W$_{2/3}$Sc$_{1/3}$)$_2$CO$_2$    &  9.90 & $-4.00$ & 38.82 & $-7.59$\\
 (W$_{2/3}$Y$_{1/3}$)$_{2}$CO$_2$    &  8.91 &$-3.36$ & 35.53 & $-4.84$\\                                                          
\hline
                                                             &   $d_{11}$  &    $d_{12}$  &   $d_{11}$   &   $d_{12}$  \\
\hline
(Mo$_{2/3}$Sc$_{1/3}$)$_2$CO$_2$  & 4.63 & $-3.02$ & 29.24 & $-14.71$\\         
(Mo$_{2/3}$Y$_{1/3}$)$_2$CO$_2$    & 11.44 & $-4.60$ & 35.91 & $-13.95$ \\ 
 (W$_{2/3}$Sc$_{1/3}$)$_{2}$CO$_{2}$    & 3.65 & $-2.29$ & 21.67& $-9.71$\\
 (W$_{2/3}$Y$_{1/3}$)$_{2}$CO$_{2}$  & 3.26 &  $-1.87$ & 24.98 & $-9.82$\\                                                               
\hline
\hline
\end{tabular}
\end{table}

\section{Conclusion}  {\label{sec:conclusion}}

The iMAX phases stand for a group of MAX phases such as (Mo$_{2/3}$Y$_{1/3}$)$_2$AlC, 
(Mo$_{2/3}$Sc$_{1/3}$)$_2$AlC, and (V$_{2/3}$Zr$_{1/3}$)$_2$AlC with in-plane ordered double 
transition metals that have recently been synthesized experimentally. Consequently, their Al exfoliated 
2D structures are named iMXenes. 
On the basis of first-principles calculations, we have studied dynamic stability and electronic structures 
of these iMAX phases as well as the corresponding iMXenes. 
We have shown that all iMAX phases are metallic, while their exfoliated iMXenes can become either 
metallic or semiconducting upon appropriate surface functionalization. We have shown that 
O-functinalized iMXenes are semiconducting and exhibit large piezoelectric coefficients. 
We have also demonstrated that some iMXenes have unique mechanical properties. 
Because of the many interesting properties expected for iMXenes, extensive experimental studies are highly 
anticipated in the near future.

\section*{Acknowledgments}
We would like to thank Dr. Wenbin Li for fruitful discussions. 
M.K. and A.R. are grateful to RIKEN Advanced Center for Computing and Communication (ACCC) for 
the allocation of computational resources of the RIKEN supercomputer 
system (HOKUSAI GreatWave). M.K. acknowledges the members of the Numerical
Materials Simulator at NIMS for their continued support of the supercomputing facility.
M.K. also gratefully acknowledges the support by a Grant-in-Aid for Scientific Research (No. 17K14804) 
from MEXT, Japan. C.S. acknowledges the support from the Scientific and Technological Research Council of 
Turkey (TUBITAK-116F080) and Anadolu University (BAP-1705F335). 
V.W. acknowledges the support by the Natural Science Basic Research Plan (Grant No. 2017JM1008) of the Shaanxi Province of China.

\end{document}